\PassOptionsToPackage{compatibility=false}{caption}
\documentclass[letterpaper, 10 pt, conference]{ieeeconf}

\IEEEoverridecommandlockouts
\usepackage[utf8]{inputenc}
\usepackage[american]{babel}
\usepackage{csquotes}
\usepackage[letterpaper,left=48pt,bottom=43pt,right=48pt,top=57pt]{geometry}

\usepackage{comment}
\usepackage{url}
\usepackage{graphicx}
\usepackage{subcaption}
\usepackage{multicol}
\usepackage{xspace}

\usepackage{amsmath, amssymb, amsfonts}
\usepackage{mathtools}
\usepackage{multirow}
\usepackage{tabularx}
\usepackage[binary-units]{siunitx}

\newcolumntype{C}{>{\centering\arraybackslash}X}
\newcolumntype{L}{>{\raggedright\arraybackslash}X}
\usepackage{siunitx}
\usepackage{listings}
\usepackage{float}

\newtheorem{definition}{Definition}

\newtheorem{remarkth}[definition]{Remark}
\newenvironment{remark}{\begin{remarkth}\upshape}{\end{remarkth}}

\newcommand{\dd}{\mathrm{d}}
\newcommand{\liedv}[1]{\mathcal{L}_{#1}}

\usepackage{tikz}
\usetikzlibrary{matrix,arrows,calc,positioning,shapes,decorations.pathreplacing, angles, quotes, patterns}
\tikzstyle{vertex}=[circle,fill=black!20,minimum size=15pt,inner sep=0pt]
\tikzstyle{selected vertex} = [vertex, fill=red!24]
\tikzstyle{edge} = [draw,thick,-]
\tikzstyle{dedge} = [draw,thick,<->]
\tikzstyle{shadowdedge} = [draw, dotted,->]
\tikzstyle{weight} = [font=\small]
\tikzstyle{selected edge} = [draw,line width=5pt,-,red!50]
\tikzstyle{ignored edge} = [draw,line width=5pt,-,black!20]

\usepackage{booktabs}

\title{\LARGE \bf Discrete Mechanics and Optimal Control for\\ Passive Walking with Foot Slippage}

\author{Alexandre Anahory Simoes$^{1}$, Asier L\'opez-Gord\'on$^{2}$, Anthony Bloch$^{3}$ and Leonardo Colombo$^{4}$%
\thanks{$^1$ A.~Anahory Simoes (alexandre.anahory@ie.edu) is with School of Sciences and Technology, IE University, P. de la Castellana 259, 28046 Madrid, Spain.}
\thanks{$^2$A.~López-Gordón~(asier.lopez@icmat.es) is with Instituto de Ciencias Matemáticas (ICMAT-CSIC), C/ Nicol\'as Cabrera, 13-15, 28049 Madrid, Spain.}
\thanks{$^3$A.~Bloch~(abloch@umich.edu) is with Department of Mathematics, University of Michigan, Ann Arbor, MI 48109, USA.}
\thanks{$^4$L.~Colombo (leonardo.colombo@car.upm-csic.es) is with Centro de Automática y Robótica (CAR-CSIC), Ctra. M300 Campo Real, Km 0, 200, Arganda del Rey - 28500 Madrid, Spain.}
\thanks{The authors acknowledge financial support from Grant PID2019-106715GB-C21 funded by MCIN/AEI/ 10.13039/501100011033. Asier López-Gordón also recieved support from the Grant CEX2019-000904-S funded by MCIN/AEI/ 10.13039/501100011033. A.L.G would also like to thank MCIN for the predoctoral contract PRE2020-093814. A.M.B. was partially supported by NSF grants DMS-1613819 and DMS-2103026, and AFOSR grant FA
9550-22-1-0215.}
}

\begin{document}
\newgeometry{left=54pt,bottom=54pt,right=54pt,top=54pt}

\maketitle
\thispagestyle{empty}
\pagestyle{empty}

\begin{abstract}
Forced variational integrators are given by the discretization of the Lagrange-d'Alembert principle for systems subject to external forces, and have proved useful for numerical simulation studies of complex dynamical systems. In this paper we model a passive walker with foot slip by using techniques of geometric mechanics, and we construct forced variational integrators for the system. Moreover, we present a methodology for generating (locally) optimal control policies for simple hybrid holonomically constrained forced Lagrangian systems, based on discrete mechanics, applied to a controlled walker with foot slip in a trajectory tracking problem. 
\end{abstract}

\section{Introduction}

Passive-dynamic walkers \cite{Raibert1986}, \cite{Collins2005}, \cite{McGeer1990},  are templates for human-like walking, describing its biomechanical and energetic aspects. They are uncontrolled and unpowered mechanisms that balance themselves as they walk, similarly to how people walk down a slope. 
They are modeled as dissipative dynamical systems since energy is lost when collisions are made with the ground. Mastering passive dynamics helps to understand the mechanics of walking. Passive dynamic walkers are piecewise holonomic systems \cite{Holmes2006a}, \cite{Ruina1998}, that is, mechanical systems that change at each transition of the dynamics, although they are holonomic within each stride before impacts occur \cite{Ruina1998}. 

An implicit assumption for passive-dynamic walkers is that the feet do not slip on contact with the ground. In this paper, we model passive walkers with foot slippage over a flat ground inspired by \cite{Clark2018}. In comparison with the model proposed by \cite{Clark2018}, our model avoids incrementing the dimension of the configuration space to include Lagrange multipliers. In our approach, we reduce the dynamics of the system to the constraint submanifold generated by the constraints associated with the proposed model for walking with foot slip, giving rise to motion equations with fewer degrees of freedom than \cite{Clark2018}. 

Trajectory optimization algorithms aim to find an input trajectory that minimizes a cost function subject to a set of constraints on the system's states and inputs. Trajectory optimization has been implemented extensively for systems with continuous-time dynamics, but many applications in control theory and robotics include impacts and friction contacts making the dynamics non-smooth. In this paper, we develop a trajectory optimization policy for a passive walker experiencing foot slip by introducing controls into the passive walker and by defining geometric integrators for a class of hybrid mechanical systems- that is,  (smooth) dynamical systems together with a discrete transition (impact map)- by using discrete (geometric) mechanics techniques. 

Variational integrators are a class of geometric integrators for Lagrangian systems derived from a discrete variational principle as discussed e.g. by \cite{Marsden2001a} and \cite{Hairer2006}. These integrators retain some of the main geometric properties of  continuous systems, such as symplecticity and momentum conservation (as long as the symmetry survives the discretization procedure), and good (bounded) behavior of the energy associated to the system. This class of numerical methods has been applied to a wide range of problems including optimal control \cite{Ober-Blobaum2011, colombo2016geometric}, constrained systems \cite{leyendecker2008variational,colombo2013higher}, power systems \cite{ober2013variational}, nonholonomic systems \cite{cortes2001non}, \cite{de2004geometric,colombo2015variational}, multi-agent systems \cite{colombo2020forced, colombo2021forced, colombo2022variational}, and systems on Lie groups \cite{kobilarov2011discrete}, \cite{jimenez2013discrete}. Variational integrators for hybrid mechanical systems were used in \cite{fetecau2003nonsmooth} and \cite{flasskamp2011variational}. However, these works do not consider the problem of trajectory generation. Such a problem is considered in \cite{Pekarek2007} but for the compass gait biped, while in this work we consider passive walkers under foot slippage.

The main contributions of this work are summarized as follows: 

\begin{itemize}
    \item We introduce simple hybrid holonomically constrained forced Lagrangian systems and we construct forced variational integrators for this class of hybrid mechanical system. 
    \item We introduce a reduced dynamical model for walking with foot slip and derive variational integrators for the proposed model.
        \item We present a methodology for generating (locally) optimal control policies for simple hybrid holonomically constrained forced Lagrangian systems and illustrate the method with a passive walker experiencing foot slip.
    \item We design discrete-time sub-optimal trajectories to reach a desired configuration via discrete mechanics and optimal control as the solution to a constrained nonlinear optimization problem.
 
\end{itemize}


The remainder of the paper is structured as follows. In Section \ref{section2} the constrained Lagrange-d'Alembert principle is used to derive forced Euler-Lagrage equations for mechanical systems subject to holonomic constraints. After defining simple hybrid holonomically
constrained forced Lagrangian system in \ref{sec3.1}, this is applied to construct a model for a passive walker with foot slip in Section \ref{sec3.2}. In Section \ref{sec: fc} we construct variational integrators for a passive walker experiencing foot slip. Finally, in Section \ref{sec5} we introduce controls into the model and we employ a variational integrator for the uncontrolled system, together with a suitable discretization of the cost function associated to an optimal control problem, to derive optimal control policies for trajectory generation in a tracking problem. 
\section{Constrained Lagrange-d'Alembert Principle}\label{section2}
 Let $Q$ be an $n$-dimensional differentiable
manifold with local coordinates $(q^A)$, $1\leq A\leq n$, the
configuration space of a mechanical system. Denote by $TQ$ its
tangent bundle with induced local coordinates $(q^A, \dot{q}^A)$.
Given a Lagrangian function $L:TQ\rightarrow \mathbb{R}$, its Euler--Lagrange
equations are
\begin{equation}\label{qwer}
\frac{\mathrm{d} } {\mathrm{d}t}\left(\frac{\partial L}{\partial\dot
q^A}\right)-\frac{\partial L}{\partial q^A}=0, \quad 1\leq A\leq n.
\end{equation}
In general, Eqs.~\eqref{qwer} determine a system of implicit second-order
differential equations. If we assume that the Lagrangian is hyper-regular,
that is, the ${n\times n}$ matrix $\mathcal{M}_{AB}=\left(\frac{\partial^{2} L}{\partial \dot q^A
\partial \dot q^B}\right)$ is non-singular, the local existence and uniqueness of solutions is guaranteed for any given initial condition.

Assume $L:TQ\to\mathbb{R}$ is a hyper-regular Lagrangian and that $q(t)$ satisfies Euler--Lagrange equations \eqref{qwer}. Then, there is a smooth Lagrangian vector field $f_{L}$ on $TQ$, associated with $L$, that is, there is a dynamical system associated to the Lagrangian. For $t\in[t_0,t_f]$ we say that $\gamma(t)=(q(t),\dot{q}(t))$ is a solution of $f_{L}$ with initial condition $\gamma(t_0)=(q(t_0),\dot{q}(t_0))$ if $\dot{\gamma}(t)=f_{L}(\gamma(t))$. The Lagrangian vector field $f_{L}$ associated with $L:TQ\to\mathbb{R}$ takes the form \begin{equation}\label{vf}f_{L}(q^{A},\dot{q}^{A})=\left(
 \dot{q}^{A},
  \mathcal{M}^{AB}\left(\frac{\partial L}{\partial q^{A}}-\frac{\partial L}{\partial\dot{q}^{A}\partial q^{B}}\dot{q}^{B}\right)\right)\end{equation} where $\mathcal{M}^{AB}$ denotes the inverse matrix of $\mathcal{M}_{AB}$.

A constrained Lagrange-d'Alembert principle (or principle of virtual work) for systems subject to external non-conservative forces and holonomic constraints (i.e., constraints of the form $\Phi(q)=0$ with $\Phi:Q\to\mathbb{R}$ a smooth function and $0$ a regular value of $\Phi$), establishes that the natural motions of the system are those paths $q:[0,T]\to Q\times\mathbb{R}^{m}$ satisfying 
\begin{equation}
\label{ldp}\delta\int_{0}^{T}\left(L(q,\dot{q})+\lambda\Phi(q)\right)\,\dd t+\int_{0}^{T}F(q,\dot{q})\delta q\,\dd t=0
\end{equation} 
for  null  boundary  variations $\delta q(0)=\delta q(T)=0$ and $\delta\lambda\in\mathbb{R}^{m}$. Here $\lambda(t)\in\mathbb{R}^{m}$ represents the vector of time-dependent Lagrange multipliers. The first term in Eq.~\eqref{ldp} is  the action variation augmented with a Lagrange multiplier to ensure the dynamics satisfy the constraint, while the second is known as virtual work since $F(q,\dot{q})\delta q$ is the virtual work done by the force field $F$ with a virtual displacement $\delta q$. Denoting the Jacobian of the constraints as $\displaystyle{G(q^{A})=\frac{\partial\Phi}{\partial q^{A}}}$, the constrained Lagrange-d'Alembert principle, leads to the constrained forced Euler--Lagrange equations 
\begin{equation}
\label{E-Leqns}\frac{\mathrm{d} } {\mathrm{d}t}\left(\frac{\partial L}{\partial\dot{q}^A}\right)-\frac{\partial L}{\partial q^A}-F(q_i,\dot{q}_i)+G^{T}(q^A)\lambda=0,
\end{equation} 
together with $\Phi(q)=0$. Note that the term $-G^{T}(q^A)\lambda\in T^{*}Q$ is the force imposing the system to remain in the constraint submanifold defined by $\Phi(q)=0$. 

Now, consider  the  augmented Lagrangian $\bar{L}:TQ\times\mathbb{R}^m\to\mathbb{R}$ given by $\bar{L}(q,\dot{q},\lambda)=L(q,\dot{q})+\lambda\Phi(q)$. If the Lagrangian $L$ is hyperregular, it  induces  a  well  defined  map,  the Lagrangian flow for the augmented Lagrangian, $F_t:TQ\times\mathbb{R}^m\to TQ\times\mathbb{R}^m$ by $F_t(q_{0},\dot{q}_{0},\lambda_0)\coloneqq (q(t),\dot{q}(t),\lambda(t))$, where $(q,\lambda)\in C^2([0,T],Q\times\mathbb{R}^m)$ is the unique solution of the constrained Euler--Lagrange equation with initial condition $(q_{0},\dot{q}_{0},\lambda_0)\in TQ\times\mathbb{R}^m$. Nevertheless, this approach requires one to introduce a new equation for the Lagrange multiplier, incrementing the computational costs of the optimal control problem we want to solve in Section \ref{sec5}. To overcome this issue, alternatively one can consider the submanifold $N = \{q\in Q \ | \  \Phi(q) = 0\}\subseteq Q$ and suppose that the force $F$ on the restriction to $N$, $F|_{N}:N\rightarrow T^{*}N$, is well-defined. Then the unconstrained Lagrange-d'Alembert principle associated with the restricted Lagrangian $L_{N}$ and the forces $F_{N}$ gives the same trajectories as the previous construction on the ambient manifold $Q$. In that sense we can define the flow for the holonomically constrained forced Lagrangian system as $f_N(q_{0},\dot{q}_{0})\coloneqq (q(t),\dot{q}(t))\in TN$,  where $q\in C^2([0,T],N)$ is the unique solution of the constrained forced Euler--Lagrange equation with initial condition $(q_{0},\dot{q}_{0})\in TN$.

\section{A Passive walker with foot slip}\label{sec:2}
Next, we will examine a simple case of a passive walker where the base is allowed to slide and we will formulate this system as a simple hybrid Lagrangian system. This model is inspired by  \cite{Clark2018}. In comparison with that model ours avoids incrementing the dimension of the configuration space to include Lagrange multipliers. In our approch we reduce the dynamics of the system to the constraint submanifold $N$.  

Before modeling of the passive walker with foot slip we introduce the basics of simple hybrid holonomic forced Lagrangian systems.

\subsection{Simple hybrid holonomically constrained forced Lagrangian systems}\label{sec3.1}
Simple hybrid systems \cite{Johnson1994a} (see also \cite{Westervelt2018a}) are characterized by the tuple $\mathbf{H}=(D, f, \mathcal{S}, \Delta)$, where $D$ is a smooth manifold, the \textit{domain}, $f$ is a smooth \textit{vector field} on $D$, $\mathcal{S}$ an embedded submanifold of $D$ with co-dimension $1$ called \textit{switching surface}, and $\Delta:\mathcal{S}\to D$ a smooth embedding called the \textit{impact map}. The submanifold $\mathcal{S}$ and the map $\Delta$ are also referred to as the \textit{guard} and \textit{reset map}, respectively, in \cite{Ames2006b}-\cite{Ames2006}

The dynamics associated with a hybrid systems corresponds to an autonomous system with impulse effects. We denote by $\Sigma_{\textbf{H}}$ the \textit{simple hybrid dynamical system} generated by $\mathbf{H}$, that is,  \begin{equation}\label{LHS}\Sigma_{\textbf{H}}:\begin{cases} \dot{x}(t)=f(x(t)),\quad\quad x^{-}(t)\notin\mathcal{S} \\ x^{+}(t)=\Delta(x^{-}(t))\quad x^-(t)\in\mathcal{S} \end{cases}\end{equation} with $x:I\subset\mathbb{R}\to D$ and $x^{-}$, $x^{+}$ the states just before and after the moments when integral curves of $f$ intersects $\mathcal{S}$. 

\begin{remark} A solution of a simple hybrid system may experience a Zeno state if infinitely many impacts occur in a finite amount of time \cite{Ames2006b}, \cite{Lygeros2003}, \cite{Or2008}, \cite{Zhang2001}. However, by considering the class of hybrid systems given by mechanical systems with impulsive effects as in \cite{Westervelt2018a}, we exclude such behavior by considering that the set of impact times is closed and discrete, meaning that there is no chatering about an impact point and therefore excluding Zeno behavior. Necessary and sufficient conditions for the existence of Zeno behavior in the class of simple hybrid Lagrangian systems have been explored in \cite{Lamperski2007} and \cite{Or2008}.\hfill$\diamond$ 
\end{remark} 

Consider $D=TQ$ and a hyper-regular Lagrangian $L:TQ\to\mathbb{R}$. Associated with the dynamics generated by $L$, there exists a Lagrangian vector field $f_{L}$ as in \eqref{vf}. Note that $\Delta:\mathcal{S}\to TQ$ is continuous. If we denote the closure of $\Delta(\mathcal{S})$ by $\overline{\Delta(\mathcal{S})}$, then we must assume $\overline{\Delta(\mathcal{S})}\cap \mathcal{S}=\emptyset$ and, therefore, an impact does not lead immediately to another impact (see Section 4.1 \cite{Westervelt2018a} for more details). 

We further assume that $\mathcal{S}\neq\emptyset$ and there exists an open subset $U\subset TQ$ and a differentiable function $h:U\to\mathbb{R}$ such that $\mathcal{S}=\{x\in U \mid h(x)= 0\}$ with $\frac{\partial h}{\partial x}(s)\neq 0$ for all $s\in\mathcal{S}$ (that is, $\mathcal{S}$ is an embedded submanifold of $TQ$ with co-dimension $1$) and the Lie derivative of the vector field $f_{L}$ with respect to $h$ does not vanish on $TQ$, that is $\liedv{f_{L}}h(w)\neq 0$, $\forall w\in TQ$.  A trajectory $\gamma:[0,T]\to TQ$ crosses the switching surface $\mathcal{S}$ at $t_{i}^{-}=\hbox{inf}\{t>0|\gamma(t)\in \mathcal{S}\}$. {We allow the trajectory $\gamma(t)$ to be continuous but nonsmooth at $t_i^{-}$. That is, the velocity before the impact $\dot{q}^{-}$ is different from the velocity $q^{+}$ after the impact at $\mathcal{S}$, namely, $\dot{q}(t_i^{-})\neq \dot{q}(t_i^{+})$.  

\begin{definition}
A simple hybrid system $\mathbf{H}=(D, f, \mathcal{S}, \Delta)$ is said to be a \textit{simple hybrid holonomically constrained forced Lagrangian system} if it is determined by $\mathbf{H}^{L_N}\coloneqq (TN, f_{N}, \mathcal{S}_
N, \Delta_N)$, where $f_{N}:TN\to T(TN)$ is the flow for the holonomically constrained forced Lagrangian system as described in Section \ref{sec3.1}, and $\mathcal{S}_N$ and $\Delta_N$ are the switching surface and impact maps as described above restricted to submanifolds $N$ and $TN$, respectively.

The \textit{simple hybrid Lagrangian dynamical system} generated by $\mathbf{H}^{L_N}$ is given by
 \begin{equation*}\Sigma_{\mathbf{H}^{L_N}}:\begin{cases} \dot{x}(t)=f_{N}(x(t)), \hbox{ if } x^{-}(t)\notin\mathcal{S}_N,\\ x^{+}(t)=\Delta_N(x^{-}(t)),\hbox{ if } x^-(t)\in\mathcal{S}_N, \end{cases}\end{equation*}where $x(t)=(q(t),\dot{q}(t))\in TN $.


\end{definition}

%

That is, a \textit{trajectory} of a simple hybrid holonomically constrained forced Lagrangian system is determined by the restricted forced Lagrangian dynamics until the instant when the state attains the switching surface $\mathcal{S}_{N}$. We refer to such an instant as the \textit{impact time}. The impact map $\Delta_N$ gives new initial conditions from which $f_{N}$ evolves until the next impact occurs. Solutions for the simple hybrid holonomically constrained forced Lagrangian system $\mathbf{H}^{L_N}$, are considered right continuous and with finite left and right limits at each impact with $\mathcal{S}_N$. 

\subsection{Modeling passive walking with foot slip}\label{sec3.2}

We model a passive walker as a two-masses inverted pendulum. The mass of the foot is denoted by $m_1$ and the hip by $m_2$. The length of the leg is given by $\ell$. The angles of the leg are restricted to $\theta\in [-a, a]\subset\mathbb{R}$ (when $\theta$ hits the boundary, $-a$, a new step is taken and $\theta$ is reset to $a$). The coordinates of the center of mass will be given by $(x, y)$ and the coordinates of the foot are $(\overline{x}, \overline{y})$ (see Figure \ref{fig:model1}).

\begin{figure}[htb!]
    \centering
    \begin{tikzpicture}
        \coordinate (bob) at (2,3);
        \coordinate (origo) at (4,0);
        \draw[thick,->] (0,0) -- (6,0) node[anchor=north west] {$x$};
        \draw[thick,->] (0,0) -- (0,4) node[anchor=south east] {$y$};
        \draw[thick,-] (4,0) -- (2,3) node [anchor=north east] {$(x,y)$} node [midway, above right] {$\ell$};
        \filldraw[blue!40!white, draw=black] (origo) circle (0.1) node[anchor=north east] {\textcolor{black}{$(\overline{x},\overline{y})$}};
        \filldraw[fill=blue!40!white, draw=black] (1.85,3) rectangle (2.15,3.3);
        \draw[thick,dashed](4,0) -- (4,3) node (mary) [anchor=north west] {};
        \pic [draw, <-, "$\theta$", angle eccentricity=1.5] {angle = mary--origo--bob};
    \end{tikzpicture}
    \caption{Leg and foot: The coordinates of the foot are given by $(\overline{x},\overline{y})$, the center of mass are $(x,y)$ and $\theta$ is the angle between the leg of length $\ell$ and the vertical axis.}
    \label{fig:model1}
\end{figure}
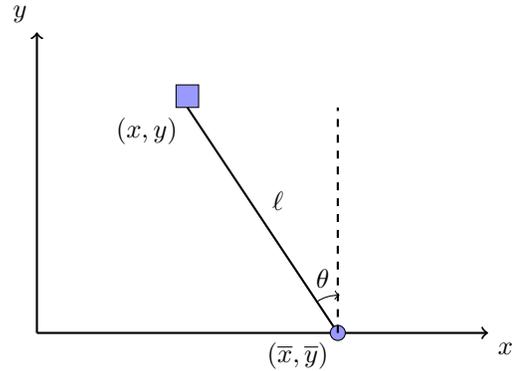

Let us denote by $m=m_1+m_2$, $I=\frac{\ell^{2}m_1m_2}{m}$, $r=\frac{\ell m_2}{m}$, where $I$ is the moment of inertia about the center of mass and $r$ is the distance from the foot to the center of mass, which is kept constant along the motion. Also, note that the coordinates of the center of the mass satisfy {$x=\overline{x} + r\sin\theta$, $y=\overline{y}+r\cos\theta$, so that the center of the mass is located along the leg at some point between the foot and the hip. In addition, we impose the constraint $\overline{y} = 0$ which means that the foot will not leave the floor. With this notation, the constraint implies that $y-r\cos\theta=0$.  

Note that for this model, $\theta$ is constrained to be in $[-a,a]$. If $\theta$ crosses the
negative boundary, we say a new step occurs and $\theta$ is reset to $a$. If $\theta$ crosses the positive boundary, specifically if $\theta = \frac{\pi}{2}$ (i.e., $x=\bar{x}$), we say that a crash has occurred. In this case, the model stops walking and we report a failure. This also implies that falling forwards is not permitted; the only way to crash is by falling backwards.

Before deriving the hybrid dynamics, we first need to determine the switching surface $\mathcal{S}$. Assume that the leg takes symmetric steps of angle $a$, that is, $\theta\in [a,-a]$. When $\theta =-a$, the angle is reset to $a$ (corresponding to a new step taking place and the swing legs switching).
Therefore, we will take the switching surface $\mathcal{S}$, to be $\mathcal{S}=\{\theta= -a\}$.

The continuous dynamics is determined by a Lagrangian function $L:TQ\to\mathbb{R}$ corresponding to a planar rigid body, where $Q=\mathbb{R}^{2}\times\mathbb{S}^{1}$ is the configuration space locally described by the coordinates $q=(x,y,\theta)$, and  $L(q,\dot{q})=K(q,\dot{q})-V(q)$, where  \begin{equation*}
K(q,\dot{q})=\frac{m}{2}(\dot{x}^{2}+\dot{y}^{2})+\frac{I\dot{\theta}^{2}}{2},\,V(q)=mgr\cos\theta,
\end{equation*}together with the (holonomic) constraint $y-r\cos\theta=0$ defining the submanifold $N$ which may be seen as diffeomorphic to $\mathbb{R}\times \mathbb{S}^{1}$. The restricted Lagrangian $L_{N}$ defined on coordinates $(x,\theta,\dot{x},\dot{\theta})$ is given by the restricted kinetic energy
$$K_{N}=\frac{m}{2}(\dot{x}^{2}+r^{2}\sin^{2} \theta \dot{\theta}^{2})+\frac{I\dot{\theta}^{2}}{2}$$
minus the restricted potential function which remains the same under the restriction to $N$.

We assume that the friction forces of the foot with the ground are non-conservative forces (conservative forces might be  included into the potential energy $V$), which are determined by a fibered map $F:TQ\to T^{*}Q$. 
The forces exerted from the friction on the foot in the configurations $q=(x,y,\theta)$ are given by 
\begin{align*}
F_x&=-\kappa\dot{\bar{x}}=-\kappa(\dot{x}+r\dot{\theta}\cos\theta),\quad F_y=0,\\
F_{\theta}&=-\kappa\dot{\bar{x}}(r\cos\theta)=-\kappa r\cos\theta(\dot{x}+r\dot{\theta}\cos\theta).
\end{align*}
This force is well-defined on the restriction to $N$. 

At a given position and velocity, the force will act against variations of the position (virtual displacements) and the dynamics should also satisfy the holonomic constraint $\Phi(q)=y-r\cos\theta=0$. 

Euler--Lagrange equations \eqref{E-Leqns} for the restricted Lagrangian $L_{N}$ and forces $F_{N}$ are given by
\begin{align}
m\ddot{x}&=-\kappa(\dot{x}+r\dot{\theta}\cos\theta) \label{eqcont4}\\
\ddot{\theta}(I+mr^{2}\sin^{2}\theta)&=-\kappa r\cos\theta(\dot{x}+r\dot{\theta}\cos\theta) \nonumber\\
&+rm\sin\theta(g-r\dot{\theta}^{2}\cos\theta) \label{eqcont5}
\end{align}
on the submanifold $N$.


The last step to describe the hybrid dynamics for the simple hybrid holonomically constrained forced Lagrangian system is to find the impact map $\Delta_N$. We assume as in \cite{Saglam2014} a rigid hip, that is, the
horizontal position and velocity of the foot do not change at impacts (see Figure \ref{fig:model2}), namely ${\bar{x}}^+={\bar{x}}^-$ and $\dot{\bar{x}}^+=\dot{\bar{x}}^-$. Additionally, we assume that the angular momentum is conserved in the impact.
 Under these assumptions (see \cite{Saglam2014,McGeer1990} for the case without foot slip and horizontal ground), the impact map is defined as the map $\Delta_N:\mathcal{S}\rightarrow TN\subseteq TQ$, where $\mathcal{S}_N=\{\theta=-a\}$, with $\Delta_N(x^{-},-a,\dot{x}^{-},\dot{\theta}^{-})=(x^{+},\theta^{+},\dot{x}^{+},\dot{\theta}^{+})$ given by
\begin{equation}
\begin{aligned}
{x^{+} - r \sin \theta^{+}}&={x^{-} - r \sin (-a)}\\
\theta^+ & = \theta^- + 2a \\
{\dot{x}^{+} - r \dot{\theta}^{+} \cos \theta^{+}}&={\dot{x}^{-} - r \dot{\theta}^{-} \cos (-a)}\\
\dot{\theta}^{+} &= \cos(2a)\dot{\theta}^{-}.
\label{impact_map_continuous}
\end{aligned}
\end{equation}

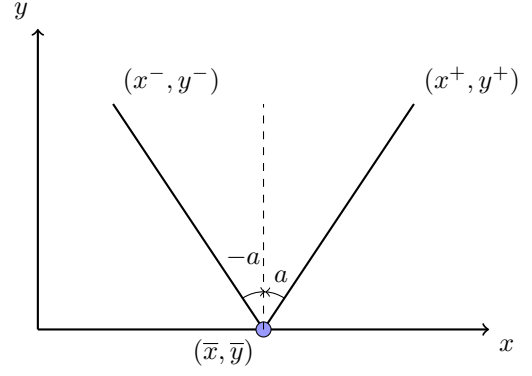
\begin{figure}[htb!]
    \centering
    \begin{tikzpicture}
        \coordinate (origo1) at (1,3);
        \coordinate (bob) at (3,0);
        \coordinate (origo2) at (5,3);

        \draw[thick,->] (0,0) -- (6,0) node[anchor=north west] {$x$};
        \draw[thick,->] (0,0) -- (0,4) node[anchor=south east] {$y$};
        \draw[thick,-] (3,0) -- (5,3) node [anchor=south west] {$(x^{+},y^{+})$};
        \draw[thick,-] (3,0) -- (1,3) node [anchor=south west] {$(x^{-},y^{-})$};
        \filldraw[blue!40!white, draw=black] (bob) circle (0.1) node[anchor=north east] {\textcolor{black}{$(\overline{x},\overline{y})$}};

        \draw[thin,dashed] (3,0) -- (3,3) node (mary) [anchor=north west] {};
        \pic [draw, <-, "$-a$", angle eccentricity=2] {angle = mary--bob--origo1};
        \pic [draw, ->, "$a$", angle eccentricity=1.5] {angle = origo2--bob--mary};
    \end{tikzpicture}
    \caption{Depiction of the impact. The position of the foot is continuous at the impact. Resetting the angle forces a reset on the position of the center of mass.}
    \label{fig:model2}
\end{figure}



\section{Forced variational integrator for a passive walker experiencing foot slip}
\label{sec: fc}
A \textit{discrete Lagrangian} is a differentiable function
$L_d\colon Q \times Q\to \mathbb{R}$, which may be considered as an
approximation of the action integral defined by a continuous regular 
Lagrangian $L\colon TQ\to \mathbb{R}.$ That is, given a time step $h>0$
small enough,
\[
L_d(q_0, q_1)\approx \int^h_0 L(q(t), \dot{q}(t))\; \dd t,
\]
where $q(t)$ is the unique solution of the Euler--Lagrange equations with  boundary conditions $q(0)=q_0$ and $q(h)=q_1$.

Construct the grid $\mathcal{T}=\{t_{k}=kh\mid k=0,\ldots,N\},$ with $Nh=T$
and define the discrete path space
$\mathcal{P}_{d}(Q)\coloneqq \{q_{d}:\{t_{k}\}_{k=0}^{N}\to Q\}.$ We
identify a discrete trajectory $q_{d}\in\mathcal{P}_{d}(Q)$ with its
image $q_{d}=\{q_{k}\}_{k=0}^{N}$, where $q_{k}\coloneqq q_{d}(t_{k})$. The
discrete action $\mathcal{A}_{d}:\mathcal{P}_{d}(Q)\to\mathbb{R}$ for this
sequence of discrete paths is calculated by summing the discrete Lagrangian on each
adjacent pair, and it is defined by \begin{equation}\label{acciondiscreta}
\mathcal{A}_d(q_{d}) = \mathcal{A}_d(q_0,...,q_N) \coloneqq \sum_{k=0}^{N-1}L_d(q_k,q_{k+1}).
\end{equation}


The discrete variational principle \cite{Marsden2001a}, states that
the solutions of the discrete system determined by $L_d$ must
extremize the action sum given fixed points $q_0$ and $q_N.$
Extremizing $\mathcal{A}_d$ over $q_k$ with $1\leq k\leq N-1,$ we
obtain the following system of difference equations
\begin{equation}\label{discreteeq}
 D_1L_d( q_k, q_{k+1})+D_2L_d( q_{k-1}, q_{k})=0.
\end{equation}
These equations are usually called \textit{the discrete Euler--Lagrange
equations}. Given a solution $\{q_{k}^{*}\}_{k\in\mathbb{N}}$ of
Eq.~\eqref{discreteeq} and assuming the regularity hypothesis, i.e., the
matrix $(D_{12}L_d(q_k, q_{k+1}))$ is regular, it is possible to
define implicitly a (local) discrete flow $
\Upsilon_{L_d}\colon\mathcal{U}_{k}\subset Q\times Q\to Q\times Q$
by $\Upsilon_{L_d}(q_{k-1}, q_k)=(q_k, q_{k+1})$ from
(\ref{discreteeq}), where $\mathcal{U}_{k}$ is a neighborhood of the
point $(q_{k-1}^{*},q_{k}^{*})$.

\subsection{Forced variational integrators for holonomically constrained forced Lagrangian systems}
The key idea of variational integrators is that the variational principle is discretized rather than the resulting equations of motion. As we explained before, we discretize the state space $TQ$ as $Q\times Q$ and consider a discrete Lagrangian $L_d:Q\times Q\to\mathbb{R}$ and, in addition, we consider discrete ``external forces'' $F_{d}^{\pm}:Q\times Q\to T^{*}Q$ approximating the continuous-time action 
and non-conservative external forces given by 
 \begin{align}
\int_{t_k}^{t_{k+1}}L(q(t),\dot{q}(t))\,dt\simeq &\,\, L^d(q_{k},q_{k+1})\label{eqq1}\\
\int_{t_k}^{t_{k+1}}F_{i}(q(t),\dot{q}(t))\delta q\,dt\simeq&F_{d}^{-}(q_{k},q_{k+1})\delta q_{k}\nonumber\\&+F_{d}^{+}(q_{k},q_{k+1})\delta q_{k+1}.\label{eqq2}
  \end{align}
Alternatively, we can directly work with a discretized version of the submanifold $N$. Here, the restricted discrete Lagrangian $L^{d}_{N}:N\times N\to\mathbb{R}$ and discrete ``external forces'' $F_{N,d}^{\pm}:N\times N\to T^{*}N$ are approximating the continuous time restricted Lagrangian and force map, respectively.

Note that, physically speaking, $F_d^{\pm}$ are not external forces. They are in fact momentum, since $F^{\pm}_{d}$ are defined by a discretization of the work done by the force $F$. The idea behind the $\pm$ is that one needs to combine the two discrete forces to give a single one-form   $F_{d}:Q\times Q\to T^{*}(Q\times Q)$ defined by $$F_{d}(q_0,q_1)(\delta q_0,\delta q_1)=F_{d}^{+}(q_0,q_1)\delta q_1+F_{d}^{-}(q_0,q_1)\delta q_0.$$

The discrete-time forced Euler--Lagrange equations on the submanifold $N$ are
 \begin{align}
0=&D_1L^{d}_{N}(q_k,q_{k+1})+D_2L^{d}_{N}(q_{k-1},q_k)\\
&+F_{N,d}^{-}(q_k,q_{k+1})+F_{N,d}^{+}(q_{k-1},q_k).\label{DEL_forced}
\end{align}

\subsection{Constrained forced variational integrator for a passive walker under foot slip}


Next, consider the midpoint (second order) discretization rule, that is, $q(t)\simeq\frac{q_k+q_{k+1}}{2}$, $\dot{q}(t)\simeq\frac{q_{k+1}-q_k}{h}$ and define the discrete Lagrangian $L_{d}:\mathbb{R}^{3}\times\mathbb{R}^{3}\to\mathbb{R}$ as $$L_d(q_k,q_{k+1})=hL\left(\frac{q_k+q_{k+1}}{2},\frac{q_{k+1}-q_k}{h}\right),$$ with $h>0$ denoting the time step and $q_k=(x_k,y_k,\theta_k)$ for $k=0,\ldots,N$.

In our model, the discrete Lagrangian $L_d:(\mathbb{R}\times\mathbb{S})\times(\mathbb{R}\times\mathbb{S})\to\mathbb{R}$ is given by
\begin{align*}
L_d=&\frac{m}{2h}(x_{k+1}-x_k)^2+\frac{1}{2h}\left(I+mr^{2}\sin^{2}\frac{\theta_{k+1}+\theta_{k}}{2}\right) \\
& \times (\theta_{k+1}-\theta_k)^2 - hmgr\cos\left(\frac{\theta_{k}+\theta_{k+1}}{2}\right).
\end{align*}


The discrete external forces are given by 
\begin{align}
  & F_d^+ = \frac{h}{2} F \left(q=\frac{q_{k-1}+q_{k}}{2}, \dot q=\frac{q_{k}-q_{k-1}}{h}  \right),\\
  &F_d^- = \frac{h}{2}F \left(q=\frac{q_{k}+q_{k+1}}{2}, \dot q=\frac{q_{k+1}-q_{k}}{h}  \right) .
\end{align}
where $F_d^+$ is evaluated in $(q_{k-1}, q_k)$ and $F_d^-$ is evaluated in $(q_{k}, q_{k+1})$.
Note that the restricted discrete force maps have the same expression.


The discrete Euler--Lagrange equations with forces are then
\begin{align*}0=&\frac{m}{h}(2x_k-x_{k+1}-x_{k-1})+hmg\sin\alpha+F_{d,x}^{-}+F_{d,x}^{+},\\
0=&\frac{I}{h}(2\theta_k-\theta_{k-1}-\theta_{k+1})\\
+&\frac{mr^{2}}{2h}\sin\frac{\theta_k+\theta_{k-1}}{2}\cos\frac{\theta_k+\theta_{k-1}}{2}(\theta_{k}-\theta_{k-1})^{2}\\
+&\frac{mr^{2}}{2h}\sin\frac{\theta_{k+1}+\theta_{k}}{2}\cos\frac{\theta_{k+1}+\theta_{k}}{2}(\theta_{k+1}-\theta_{k})^{2}+ F_{d,\theta}^{+}\\+&F^{-}_{d,\theta}
-\frac{mghr}{2}\left(\sin(\frac{\theta_k+\theta_{k-1}}{2})-\sin(\alpha-\frac{\theta_k+\theta_{k+1}}{2})\right).
\end{align*}


\section{Discrete Mechanics and optimal control for a controlled walker under foot slip}\label{sec5}


Next, we add control forces to the previous formalism. The equations of motion are now given by
\begin{equation}\label{controleq}\frac{\mathrm{d} } {\mathrm{d}t}\left(\frac{\partial L_N}{\partial\dot{q}^{A}}\right) -\frac{\partial L_N}{\partial q^{A}}=u_a\overline{Y}_{A}^{a}+(F_N)_A,\end{equation}
where $\overline{Y}^{a}=\overline{Y}^{a}_{A}(q)dq^{A},$ $1\leq a\leq
m<n$ are the control forces, $u(t)=(u_1(t),...,u_m(t))\in U$ are the control inputs, and $U$ is an open subset of $\mathbb{R}^{m}$, the set of admissible controls.
Note the previous equations give a model of an affine control system of the form 
\begin{equation}\label{eqq1}
\ddot{q}=f_{N}(q,\dot{q})+g(q,\dot{q})u,
\end{equation} where $g=(\mathcal{M}^{AB}P,0_{(n-a)\times(n-a)})^{T}$ and $P$ is a matrix mapping $u$ to the system's generalized forces. 

In a typical optimal control problem, one whishes to find a trajectory and a control minimizing a cost function of the form
\begin{equation*}
    \mathcal{J}(q,u)=\int_{0}^{T} C(q(t),\dot{q}(t),u(t)) \ \dd t
\end{equation*}
verifying a control equation such as \eqref{controleq} and, in addition, some boundary conditions giving information about the initial and terminal states of the system.

Let us suppose now that the control force is given by $\bar{Y}^{1}=\dd x$ and $\bar{Y}^{2}=\dd \theta$. Hence, we have the following controlled equations of motion on $N$
\begin{align}
&m\ddot{x}=-\kappa(\dot{x}+r\dot{\theta}\cos\theta) + u_{x},\label{eqcont6}\\
mr^{2}(\ddot{\theta}\sin^{2}\theta&+\dot{\theta}^{2}\cos\theta\sin\theta) +   I\ddot{\theta} \nonumber\\
&=-\kappa r\cos\theta  (\dot{x}+r\dot{\theta}\cos\theta) + u_{\theta}, \label{eqcont7}
\end{align}
as long as $\theta \neq -a$.

\begin{remark}
Note that in the restricted configuration space $\mathbb{R}\times \mathbb{S}^{1}$ the system is fully actuated but in the ambient space $\mathbb{R}^{2}\times \mathbb{S}^{1}$ the system is underactuated.\hfill$\diamond$
\end{remark}

Suppose that we would like to follow a known reference trajectory $\gamma_{r}:[0,T]\rightarrow Q$ denoted by
$\gamma_{r}(t)=(x_{r}(t),\theta_{r}(t))$.
We want to find a control strategy minimizing the cost functional
\begin{equation*}
    \mathcal{J}(q,u)=\frac{1}{2} \int  \varepsilon \|u\|^{2} + \eta \|\gamma_{r}-\gamma\|^{2} + \rho \|\dot{\gamma}_{r}-\dot{\gamma}\|^{2} \ \mathrm d t,
\end{equation*}
with $\gamma$ satisfying the control equations \eqref{eqcont6} and \eqref{eqcont7}.
The parameters $\varepsilon,\ \eta$ and $\rho$ are the weights of the control inputs, the trajectory-tracking and the velocity-tracking terms, respectively, 


We may transpose the optimal control problem to a nonlinear constrained optimization problem using a discretization of the principle above. Indeed, after applying the discretization procedure we come down to the problem of minimizing
\begin{equation*}
    J_{d}(q_{d},u_{d})=\sum_{k=0}^{N-1} C_{d}(q_{k},q_{k+1},u_{k},u_{k+1}),
\end{equation*}
subject to the discrete dynamics
\begin{equation}
\begin{aligned}
0=&D_1L^{d}_{N}(q_k,q_{k+1})+D_2L^{d}_{N}(q_{k-1},q_k)\\
&+F_{N,d}^{-}(q_k,q_{k+1})+F_{N,d}^{+}(q_{k-1},q_k) 
{
+ u_{k-1} + u_{k},}
\label{DEL_controls}
\end{aligned}
\end{equation}
with the boundary values $q_{0}, q_{N}$ given. Notice that Eq.~\eqref{DEL_controls} correspond to the forced discrete Euler--Lagrange equations \eqref{DEL_forced}
with $F_{N,d,u}^{\pm}(q_k,q_{k+1},u_{k}) = F_{N,d}^{\pm}(q_k,q_{k+1})+ u_{k}$.

Next, we discretize the optimal control problem. Fixing a time step $h>0$, we discretize the cost function so that
$$C_{d}(q_{k},q_{k+1},u_{k},u_{k+1}) \approx \int_{kh}^{(k+1)h} C\left(q(t), \dot{q}(t), u (t)\right) \ \dd t.$$
Thus we set
{
\begin{equation*}
        C_{d}(q_{k},q_{k+1},  u_{k},u_{k+1}) = 
        h C\left(\frac{q_{k}+q_{k+1}}{2}, \frac{q_{k+1}-q_{k}}{h}, u_{k}\right),
\end{equation*}
where $u_k = u \left(\frac{t_k+t_{k+1}}{2}\right)$.
}

The problem is subjected to the discrete dynamics
\begin{equation}\label{controlled:walker}
\begin{split}
    0=&\frac{m}{h}(2x_k-x_{k+1}-x_{k-1})+F_{d,x}^{-}+F_{d,x}^{+} \\
&+u_{x, k} + u_{x, k-1},\\
0=& m r \left\{2 g h \left(\sin \left(\frac{\theta_{k-1}+\theta_{k}}{2}\right)+\sin \left(\frac{\theta_{k}+\theta_{k+1}}{2}\right)\right)
\right .\\ & \left.
+\frac{r}{h} \left[(\theta_{k-1}-\theta_{k})^2 \sin (\theta_{k-1}+\theta_{k})
\right. \right. \\ & \left. \left.
+2 (\theta_{k-1}-\theta_{k}) \cos (\theta_{k-1}+\theta_{k})
\right. \right. \\ & \left. \left.
+(\theta_{k}-\theta_{k+1})^2 \sin (\theta_{k}+\theta_{k+1})
\right. \right. \\ & \left. \left.
+2 (\theta_{k+1}-\theta_{k}) \cos (\theta_{k}+\theta_{k+1})\right]\right\}
\\ &
- \frac{2}{h} (\theta_{k-1}-2 \theta_{k}+\theta_{k+1}) \left(2 I+m r^2\right)
\\ &
+4(F_{d,\theta}^{-}+F_{d,\theta}^{+}+u_{\theta, k-1}+u_{\theta, k}),
\end{split}
\end{equation}
and to the boundary conditions $q_{0}=(x_{0},\theta_{0})$ and $q_{N}=(x_{N},\theta_{N})$ fixed. In addition, we have the following conditions 
on the initial and final velocities:
\begin{equation}\label{DMOC:boundary:conditions}
    \begin{split}
        & \mathbb{F} L(q_{0},\dot{q}_{0})=\mathbb{F}^{F_{N,d,u}^{-}}L_{d}(q_{0},q_{1},u_{0}),\\
        & \mathbb{F} L(q_{N},\dot{q}_{N})=\mathbb{F}^{F_{N,d,u}^{+}}L_{d}(q_{N-1},q_{N-1},u_{N-1}),
    \end{split}
\end{equation}
where $\mathbb{F} L$ denotes the continuous Legendre transform and $\mathbb{F}^{F_{N,d,u}^{\pm}}$ denotes the forced discrete Legendre transform (see \cite{Marsden2001a} for instance), i.e.,
\begin{align}
  & D_2 L(q_0, \dot q_0) + D_1 L_d (q_0, q_1) + F_d^- (q_0, q_1) + u_0 = 0, \\
  & D_2 L(q_N, \dot q_N) - D_2 L_d (q_{N-1}, q_N) \nonumber \\ 
  & \qquad - F_d^+ (q_{N-1}, q_N)  + u_{N-1} = 0.
\end{align}

\begin{remark}
If we discretize the reference trajectory by evaluating it at discrete time
{$\gamma_{r}\left(h\frac{2k+1}{2}\right)=\left(x_{r}\left(h\frac{2k+1}{2}\right),\theta_{r}\left(h\frac{2k+1}{2}\right)\right)\equiv (x_{r,k},\theta_{r,k})$,}
then the { midpoint} discrete cost function reads
{
\begin{equation*}
\begin{split}
    C_{d}(q_{k}, & q_{k+1}, u_{k},u_{k+1}) = \frac{h}{2}\left[
     {\varepsilon  u_{k}^{2} }
     + \eta \left(\frac{x_{k+1}+x_{k}}{2} - x_{r,k}\right)^{2}  
     \right. \\ & \left. 
     +\eta \left(\frac{\theta_{k+1}+\theta_{k}}{2}-\theta_{r,k}\right)^{2}  
     +\rho \left(\frac{x_{k+1}-x_{k}}{h}-\dot{x}_{r,k}\right)^{2} 
     \right. \\ & \left. 
     + \rho\left(\frac{\theta_{k+1}-\theta_{k}}{h}-\dot{\theta}_{r,k}\right)^{2}\right].
\end{split}
\end{equation*}
}
\end{remark}

The discrete optimal control problem consists on finding a discrete trajectory $\{(x_{k},\theta_{k},u_{k})\}$ solution of the problem
\begin{equation}
    \begin{split}
        \min & \sum_{k=0}^{N-1}C_{d}(q_{k}, q_{k+1}, u_{k},u_{k+1}) \\
        & \text{discrete equations \eqref{controlled:walker}} \\
        & \text{boundary conditions \eqref{DMOC:boundary:conditions}}
    \end{split}
\end{equation}



Next, we incorporate impacts in the variational setting by finding a discretization of the impact set $S_{d}\subseteq Q\times Q$ and of the impact map $\Delta_{d}:S_{d} \rightarrow Q\times Q$. Let
$$S_{d}= \{ (x_{0},\theta_{0}, x_{1}, \theta_{1}) | \theta_{0}=a \}$$
and $\Delta_{d}(x_{0}^{-},-a, x_{1}^{-}, \theta_{1}^{-})$ 
is given by the discretization of Eqs.~\eqref{impact_map_continuous} via the midpoint rule:
\begin{equation*}
    \begin{split}
        & {\frac{x_{0}^{+}+x_{1}^{+}}{2} = \frac{x_{0}^{-}+x_{1}^{-}}{2} - r \sin(-a) + r \sin\left(\frac{\theta_0^{+}+\theta_1^{+}}{2}\right),} \\
        & \theta_0^{+} = 2a + \theta_0^- , \\
        & { x_{1}^{+}-x_{0}^{+} - r (\theta_{1}^{+}-\theta_{0}^{+})\cos \left(\frac{\theta_0^+ +\theta_1^+}{2}  \right) = x_{1}^{-}-x_{0}^{-} 
        }\\ & {
        - r (\theta_{1}^{-}-\theta_{0}^{-})\cos(-a),} \\
        & \theta_{1}^{+}-\theta_{0}^{+} = \cos(2a) (\theta_{1}^{-}-\theta_{0}^{-}),
    \end{split}
\end{equation*}
that is,
{
\begin{align*}
x_0^+ =&\ x_0^- -\frac{1}{2} r (\theta_0^--\theta_1^-) (\cos a -\cos (2 a) \cos \psi )\\
    &+r (\sin a  +\sin \psi ),\\
 x_1^+ =&\ x_1^- + \frac{1}{2} r (\theta_0^--\theta_1^-) (\cos a -\cos (2 a) \cos \psi ) \\
    &+r (\sin a  +\sin \psi ),\\
 \theta_0^{+} =&\ 2a + \theta_0^- , \\
  \theta_1^+ =&\  \cos (2a) (\theta_{1}^{-}-\theta_{0}^{-}) + a,
\end{align*}
where $\psi=a+\frac{1}{2}\cos (2a) (\theta_1^--\theta_0^-)$.
}

Note that the energy of the system is not preserved between impacts. Indeed, 
$$E_L=\frac{m}{2}(\dot{x}^{2}+r^{2}\sin^{2} \theta \dot{\theta}^{2})+\frac{I\dot{\theta}^{2}}{2}+mgr \cos \theta,$$
and then, 
$$E_L\circ \Delta=\cdots +\frac{I(\dot{\theta}^{+})^2}{2}
= \cdots + \frac{I\dot{\theta}^{2}}{2} \cos^2 (2a) \neq E_L.$$

We have performed a Python numerical simulation with $N=80$ steps, time step $h=0.1$, parameters $g = 9.8,\ \alpha = 0,\ \kappa = 0.2,\ r = 1,\ m = 1,\ I = 0.5,\ a=\frac{\pi}{6},\ \varepsilon=0.1,\ \eta =100,\ \rho=1$; initial values $x_0=0,\ \theta_0 =\frac{\pi}{6},\ \dot x_0 =1,$ and $\dot \theta_0 = 0.1$. The reference trajectory is given by $\gamma_r(t) = \left(\bar x_r(t) + r \cos(\theta_r(t)),\theta_r(t)\right)$ for $t_{i-1}<t<t_i$, where $\bar x_r(t)=\bar x_{r,i-1} + \dot{\bar{x}}_{r,i-1} t$ and $\theta_r(t)= a + \dot \theta_{r,i-1} (t-t_{i-1})$. The values of the parameters are $t_0=0,\ \bar x_{r,0}=0,\ \theta_{r,0}=a,\ \dot {\bar {x}}_{r,0}=1,\ \dot \theta_{r,0}=-0.08$ and $t_i$ for $i\geq 1$ is the instant of the $i$-th impact (determined by the equation $\theta(t_i)=-a$). The parameters $x_{r,i},\ \theta_{r,i},\ \dot x_{r,i},\ \dot \theta_{r,i}$ are defined by the impact map \eqref{impact_map_continuous}.

The evolution of the $x$- and $\theta$-coordinates of the center of mass are plotted in Figs.~\ref{fig:simulationx} and \ref{fig:simulationtheta}, respectively; comparing
them with the reference trajectory. The curves that the center of mass, the foot, the leg and the reference trajectory describe on the $xy$-plane are represented in Fig.~\ref{fig:simulationxy}.
One can clearly observe how the trajectory of the foot approaches the reference one. The evolution of the control inputs is represented in Fig.~\ref{fig:simulationcontrols}. 

\begin{figure}[htb!]
    \centering
    \includegraphics[width=\linewidth]{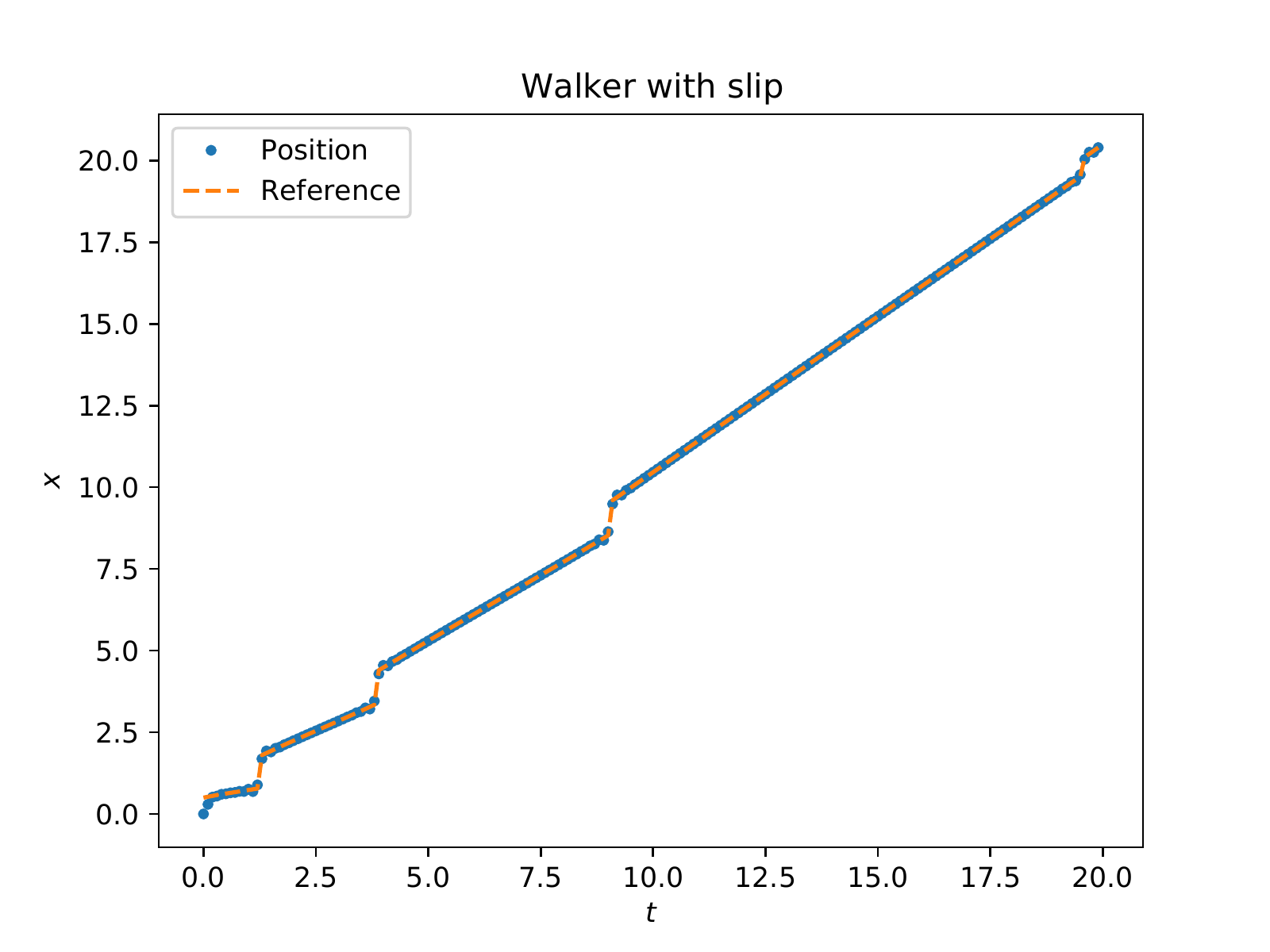}
    \caption{Horizontal components of the position of the center of mass and the reference trajectory as functions of time. }
    \label{fig:simulationx}
\end{figure}

\begin{figure}[htb!]
    \centering
    \includegraphics[width=\linewidth]{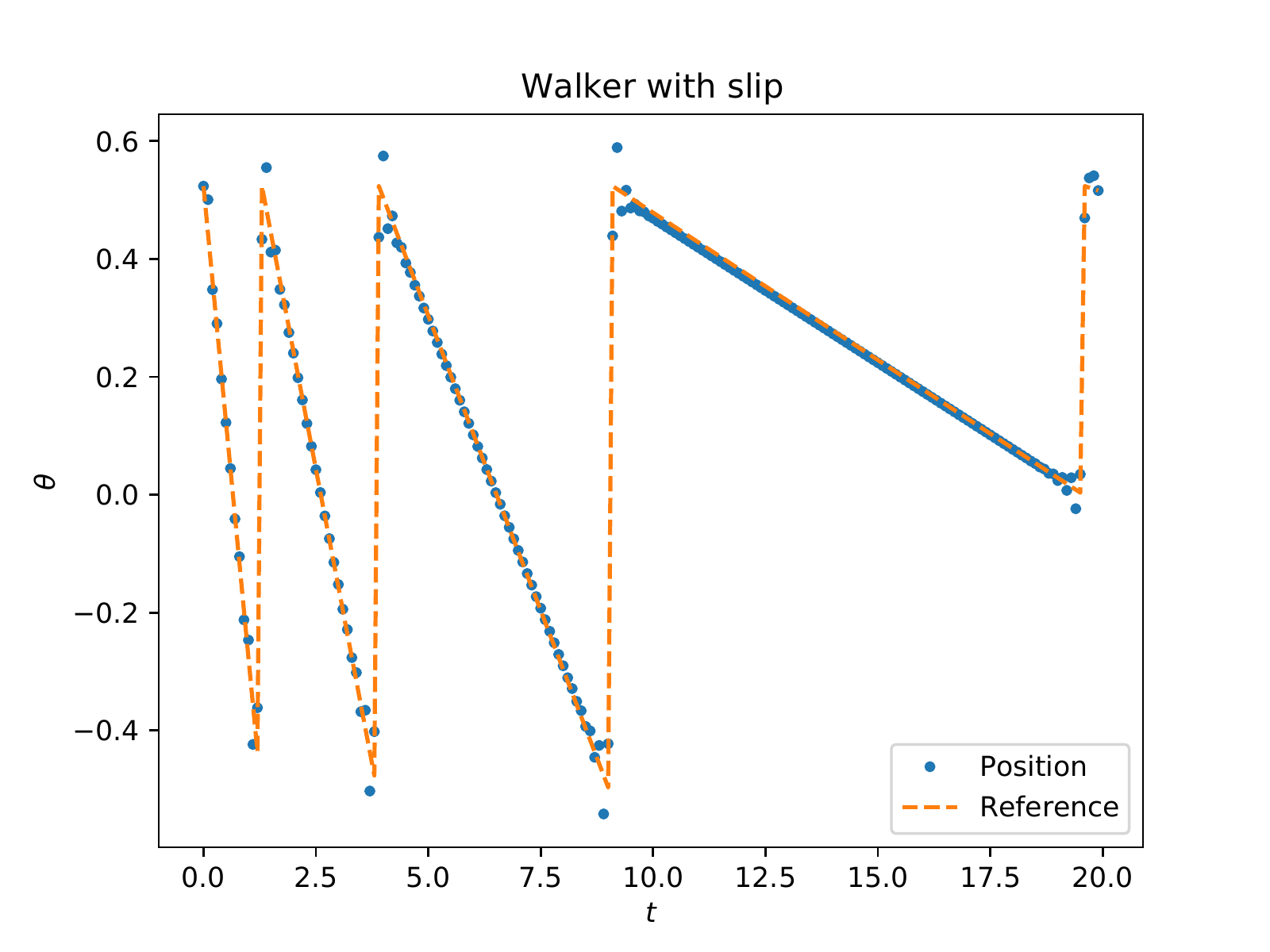}
    \caption{Angular components of the position of the center of mass and the reference trajectory as functions of time.}
    \label{fig:simulationtheta}
\end{figure}

\begin{figure}[htb!]
    \centering
    \includegraphics[width=\linewidth]{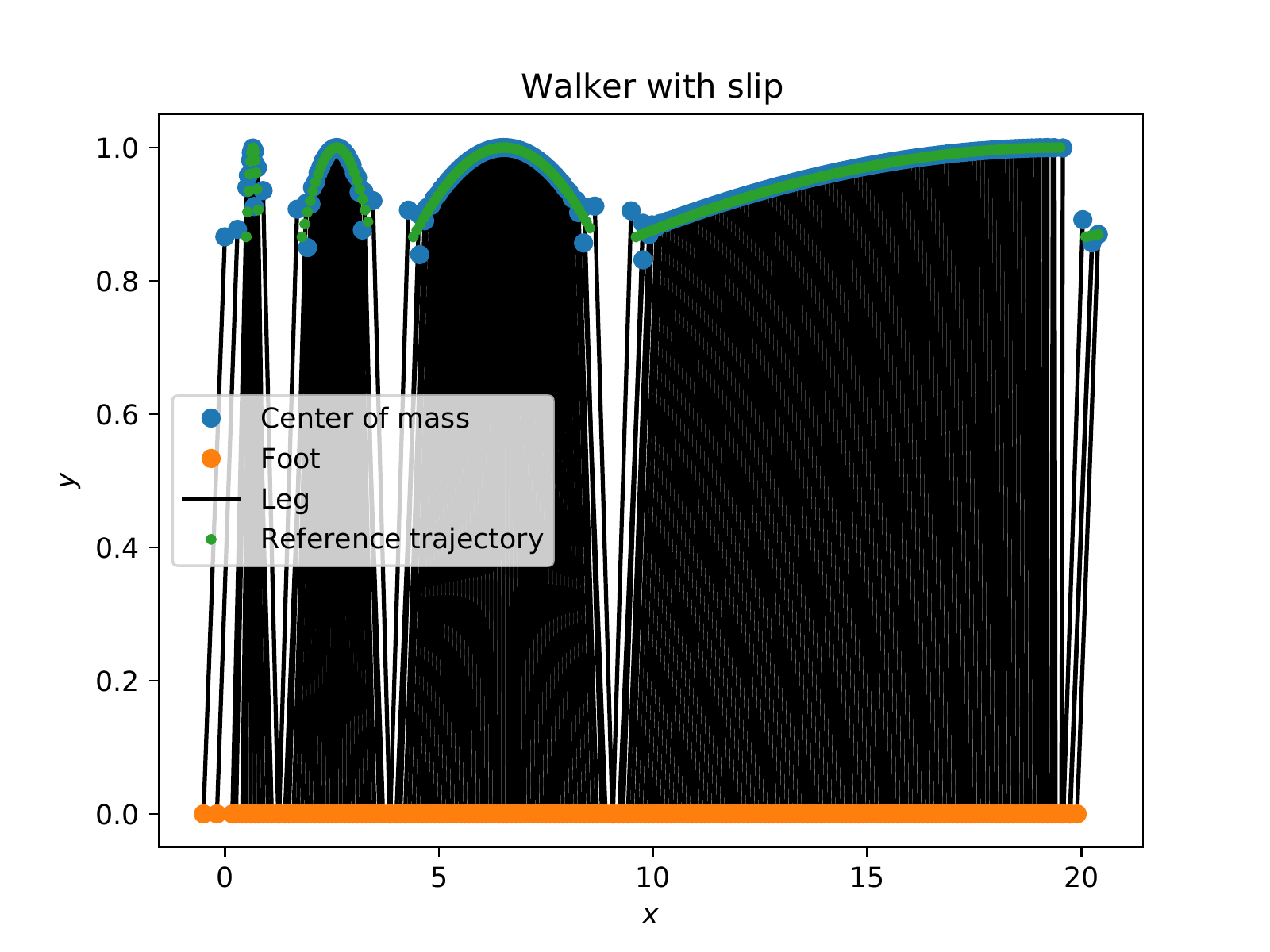}
    \caption{Trajectories of the center of mass and the foot compared with the reference trajectory.}
    \label{fig:simulationxy}
\end{figure}

\begin{figure}[htb!]
    \centering
    \includegraphics[width=\linewidth]{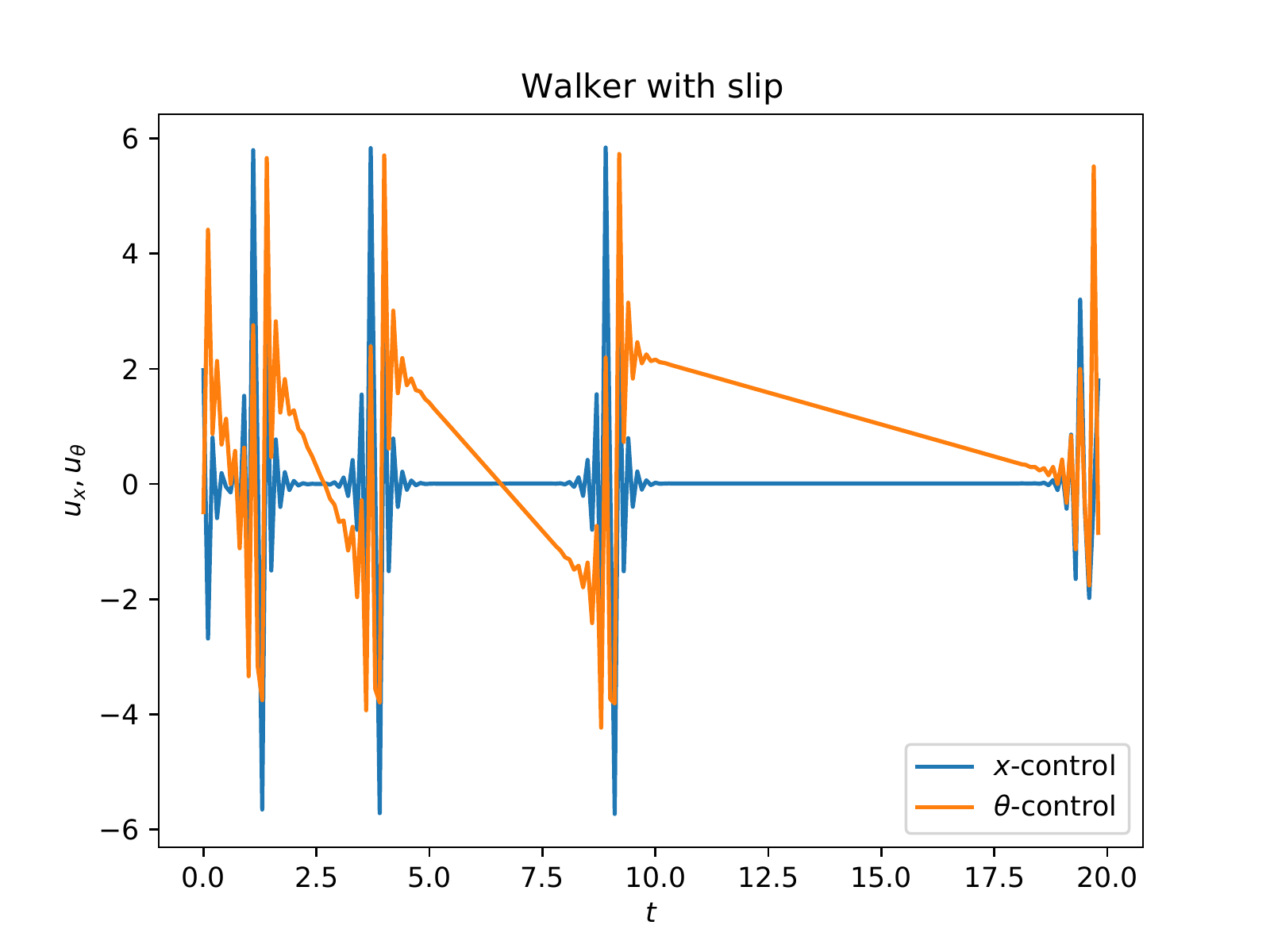}
    \caption{Horizontal and angular components of the control inputs as functions of time.}
    \label{fig:simulationcontrols}
\end{figure}




\section{Conclusions}
We have introduced simple hybrid holonomically constrained
forced Lagrangian systems and we have constructed forced variational integrators for this class of hybrid system subject to holonomic constraints. In particular, we applied the discretization to a model of a passive walker with foot slip. This  discretization is employed in a trajectory generation problem, together with a suitable discretization of a cost function, in a trajectory tracking task. This study sheds light on how to identify in the model when the walker falls. It also indicates how to design controllers based on momentum balance \cite{bayon2020can} in order to avoid falls while  tracking where leg amplitudes are equal. 
\nocite{*}

\bibliography{biblio}
\bibliographystyle{IEEEtran}

\clearpage 

\end{document}